\documentclass[twocolumn,10pt]{asme2ej}

\usepackage{graphicx} 
\usepackage{hyperref}   
\hypersetup{
	colorlinks=true,
	linkcolor=blue,
	citecolor=blue,
	urlcolor=blue,
}
\usepackage[square,numbers]{natbib}

\usepackage{amssymb,amsmath}
\usepackage{unit}
\usepackage[linesnumbered,ruled,vlined]{algorithm2e}
\newcommand{\figref}[1]{Fig.~\ref{#1}}
\def\eqref#1{Eq.~(\ref{#1})}
\def\secref#1{Sec.~\ref{#1}}

\title{Data-driven Exponential Framing for Pulsive Temporal Patterns without Repetition or Singularity}

\author{Yohei Kono
    \affiliation{
	Chief Researcher\\
	Research \& Development Group\\
	Hitachi, Ltd.\\
	Yoshidacho 292, Totsuka, Yokohama, 244-0817\\
    Japan\\
    Email: yohei.kono.un@hitachi.com
    }	
}

\author{Yoshiyuki Tajima
    \affiliation{
	Manager\\
	Research \& Development Group\\
	Hitachi, Ltd.\\
	Yoshidacho 292, Totsuka, Yokohama, 244-0817\\
    Japan\\
    Email: yoshiyuki.tajima.hh@hitachi.com
    }	
}

\begin{document}

\maketitle    

\begin{abstract}
{\it Extracting pulsive temporal patterns from a small dataset without their repetition or singularity shows significant importance in manufacturing applications but does not sufficiently attract scientific attention.
We propose to quantify how long temporal patterns appear without relying on their repetition or singularity,
enabling to extract such temporal patterns from a small dataset.
Inspired by the celebrated time delay embedding and data-driven Hankel matrix analysis,
we introduce a linear dynamical system model on the time-delay coordinates behind the data
to derive the discrete-time bases each of which has a distinct exponential decay constant.
The derived bases are fitted onto subsequences that are extracted with a sliding window in order to quantify how long patterns are dominant in the set of subsequences.
We call the quantification method Data-driven Exponential Framing (DEF).
A toy model-based experiment shows that DEF can identify multiple patterns with distinct lengths.
DEF is also applied to electric current measurement on a punching machine, showing its possibility to extract multiple patterns from real-world oscillatory data.
}
\end{abstract}


\section{Introduction}
\label{sec:intro}

Temporal patterns, meaning subsequences which are in some sense representative of their original time-series data,
play a crucial role in various research areas \cite{Scafetta2002,Leo2016,Keogh2006,Gugulothu2018,Fu2001,Lovric2014}.
This paper suggests an alternate perspective on continuous-valued temporal patterns that is more suitable for manufacturing applications.

In manufacturing processes, the development of sensing and information technologies has accelerated data-driven enhancement of their operation and maintenance.
This trend motivates the use of temporal patterns for monitoring complex phenomena that cannot be described in closed-form models.
For instance, in the case of injection-molding machines, Takatsugi, et al. \cite{Takatsugi2017} reported that temporal patterns in load torque data can characterize in-situ resin pressure,
and proposed detecting the reverse flow of resin by monitoring the data.
Similarly, for punching machines, Takahashi, et al. \cite{Takahashi2013} analyzed a temporal pattern in die acceleration data to monitor the tool wear.
For human manufacturing, Terada, et al. \cite{Terada2021} developed an unsupervised method for detecting a series of repetitive actions by recognizing basic actions from human motion data.
We also developed an anomaly detection method based on temporal patterns and applied it to manufacturing machines \cite{Kono2022}.

Motivated by the above studies,
we address data-driven extraction of temporal patterns specific to manufacturing processes.
The time-series data and patterns addressed here are characterized by the three assumptions.
Firstly, as in manufacturing machines operated with the position or speed control of their tools (e.g., Ref.~\cite{Ulsoy1993}) and chemical processes controlled on different levels of temperature reference,
reference values (setpoints) for control of manufacturing processes can jump discretely,
leading to process data containing rapidly rising (or falling) curves or irregular oscillations.
These patterns are referred to as hereafter {\em pulsive temporal patterns} in this paper.
Secondly, the manufacturing process has multiple modes of operation where the frequency of occurrence of each mode is not necessarily high.
This means that the number of occurrence of each temporal pattern can be few in number, including one-shot occurrence.
Thirdly, pulsive temporal patterns relevant in engineering practice are not restricted to anomalous events or singular behaviors.
Many relevant cases, such as transient responses to external inputs or one-off operational changes, occur without singularities, implying the necessity to develop a framework that can address one-shot patterns,
with or without singularity, beyond the conventional scope of anomaly detection.
Therefore, this paper aims to extract pulsive temporal patterns from a small dataset that may lack repeated instances.

Research on extracting pulsive temporal patterns has largely been conducted using machine learning techniques, specifically {\em time-series data mining} \cite{Esling2012}.
Different definitions of temporal patterns exist based on the machine learning task at hand.
Ref.~\cite{Lin2002} defined previously unknown, frequently occurring patterns as \emph{motifs} and developed a discovery algorithm for them.
In the last two decades, many research groups have developed sophisticated algorithms for finding motifs \cite{Lam2011,Imamura2021}.
Ref.~\cite{Keogh2005} introduced time series \emph{discords}, which are subsequences of a longer time series that are maximally different from all the rest,
making it highly compatible with anomaly detection;
see, e.g., \cite{Mueen2010,Imamura2015}.
Ref.~\cite{Ye2009} addressed the classification problem of time series and formulated a concept of {\em time series shapelet}, which are subsequences maximally representative of a class.
Jointly learning shapelets and a classifier has also been reported in Refs.~\cite{Yamaguchi2019,Yamaguchi2020}.
These studies suggest that previous definitions of temporal patterns rely on repetition or singularity.

However, to the best of our knowledge, not enough attention has been given to identifying pulsive patterns in small datasets.
This suggests that additional criteria, beyond repetition or singularity, are necessary to characterize temporal patterns.

To quantify how long patterns appear in pulsive time series in an unsupervised manner,
we propose an alternative view of the temporal patterns based on their length.
Inspired by the celebrated time delay embedding \cite{Takens1981,Abarbanel1994,Tan2023} and data-driven Hankel matrix analysis \cite{Hassani2007,LeClainche2017,Brunton2017},
we introduce a linear dynamical system model on the time-delay coordinates behind the data to derive the discrete-time bases, each of which has a distinct exponential decay constant.
These bases are then fitted onto subsequences that are extracted with a sliding window to determine how long patterns are dominant in the set of subsequences.
We call this quantification method \emph{Data-driven Exponential Framing} (DEF).
We demonstrate the effectiveness of DEF using artificial data generated by a toy model based on the harmonic oscillator.
Our results show that DEF can identify multiple patterns with distinct lengths.
In addition, we apply DEF to electric current data measured in a punching machine and show its effectiveness in this real-world scenario.
Note that this paper is a joint submission with our conference paper \cite{Kono_MECC2025},
and the discrete-time bases with exponential decay constants have been proposed in our non-archival conference paper \cite{Kono2022}.

The contributions of this paper are twofold.
The first capability is the development of a practical method for characterizing small timeseries datasets.
The method extracts dominant temporal patterns from a timeseries dataset without their repetition or singularity,
which can be search queries for the timeseries data mining.
The second contribution is the construction of an alternate framework for decomposing timeseries based on the exponential decay rate.
This DEF framework enables the identification of multiple (complex) frequencies that explain pulsive patterns in timeseries data.
In simpler terms, DEF offers an alternative interpretation of the modal components calculated by the Singular Spectrum Analysis \cite{Hassani2007,Golyandina2001} or the extensions of Dynamic Mode Decomposition (DMD) \cite{Schmid2010} to Hankel matrices \cite{LeClainche2017,Brunton2017}.

The rest of this paper is organized as follows.
\secref{sec:proposed} focuses on explaining the mathematical concept of DEF and its computation algorithm.
In \secref{sec:numerical}, we provide evidence that DEF can identify the lengths of dominant subsequences in numerical data obtained from a toy model.
Additionally, in \secref{sec:demo}, we demonstrate that DEF can identify multiple patterns with different lengths that are inherent in time series obtained from a manufacturing machine.
\secref{sec:ssa_dmd} shows comparative analysis with SSA and DMD, which evaluates the novelty of proposed DEF method.
Finally, \secref{sec:concusions} concludes this paper for future research.

\section{Proposed method}
\label{sec:proposed}

We begin by formulating a state-space description for a scalar output $y(t)$.
Let $\vct{x}(t)\in\mathbb{R}^n$ and $\vct{u}(t)\in\mathbb{R}^m$ be the state and exogenous input at time $t$.
The state-space systems are formulated as
\begin{align}
  \frac{\dd \vct{x}}{\dd t} &= f(\vct{x}, \vct{u}),\label{eq:dyn}\\
  y &= g(\vct{x}),\label{eq:obs}
\end{align}
where $f: \mathbb{R}^n\times \mathbb{R}^m \to \mathbb{R}^n$ is the map of dynamics and $g: \mathbb{R}^n \to \mathbb{R}$ the observable.
We consider \eqref{eq:dyn} as closed-loop (controlled) systems to represent both the physical properties of manufacturing machines and their controllers' properties.
This means that the $\vct{u}(t)$ can include reference signals (setpoints) for controlling manufacturing machines as well as control parameters.
As in a wide range of industrial machines that have a finite number of operation modes,
we assume that the reference signals and control parameters jump only under certain conditions, and otherwise remain constant.
Therefore, $\vct{u}(t)$ can be represented by a piecewise constant PWC function, which is a combination of finite-length step signals.
As a result, $\{y(t)\}$ involves rapidly rising (or falling) curves or irregular oscillations near the jump, referred to as pulsive patterns.
When exogenous inputs are not PWC, the formulation below extends in a straightforward manner by augmenting the delay vector with a finite number of input-lag terms; 
we keep the exposition focused on the PWC case for simplicity, and discuss this extension in Appendix A.

\subsection{Mathematical idea}
\label{sec:idea}

We propose a method for quantifying the duration of dominant patterns in a given time series.
The approach involves analyzing the decaying components within the time series and estimating them based on the evolution of subsequences.
This allows us to better understand the behavior of the time series and identify the most significant patterns that contribute to its overall trend.

In order to analyze the patterns in $y(t)$, we first introduce the dynamical formulation of subsequences.
Since our focus is on characterizing patterns included in $y(t)$,
we rewrite \eqref{eq:dyn} as the {\em piecewise autonomous} system by viewing $\vct{u}(t)$ as a certain fixed function.
We also assume that the state $\vct{x}$ is kept near its equilibrium by control efforts,
which allows us to linearize the nonlinear system $\{f, g\}$.
Inspired by the delay embedding techniques (see, e.g., \cite{Takens1981})
and their capability of formulating state-space dynamics \cite{Phan1998},
we consider identifying the target linear dynamics from the data on $y(t)$:
\begin{align}
  \frac{\dd\vct{x}_d}{\dd t} = {\sf A}_d \vct{x}_d(t),
  \label{eq:delay_model}
\end{align}
where $\vct{x}_d \in \mathbb{R}^d$ is the delay-coordinate vector with the degree $d$,
and written with the sampling period $\Delta t$:
\begin{align}
  \vct{x}_d(t) = \left[
  \begin{array}{c}
  y(t)\\
  y(t-\Delta t)\\
  \vdots\\
  y(t-(d-1)\Delta t)
  \end{array}
  \right].
 \end{align}
Note that \eqref{eq:delay_model} models the time-evolution of subsequences.
We will use this perspective to decompose time series data by their exponential decay rates.

Let us develop a method for characterizing timescales contained in the subsequences.
Assuming that ${\sf A}_d$ is a full-rank matrix,
it can be factorized with $d$ linearly independent eigenvectors $\vct{V}_1 \cdots \vct{V}_d \in\mathbb{C}^d$ and the corresponding eigenvalues $\lambda_1, \lambda_2, \ldots, \lambda_d$: 
\begin{align}
{\sf A}_d = {\sf V}_d {\sf \Lambda}_d {\sf V}_d^{-1},
\label{eq:eigendecomposition}
\end{align}
where ${\sf V}_d = \left[\vct{V}_1 \cdots \vct{V}_d\right]$ and ${\sf \Lambda}_d$ is the diagonal matrix whose diagonal elements are $\lambda_1, \lambda_2, \ldots, \lambda_d$.
Note that the $\vct{V}_1, \cdots, \vct{V}_d$ are equivalent to the modal components calculated by SSA \cite{Hassani2007} and DMD with the Hankel matrices \cite{LeClainche2017,Brunton2017}.
Based on this decomposition, we consider predicting $\vct{x}_d$ from an initial time $t=t_0$.
Combining \eqref{eq:delay_model} and \eqref{eq:eigendecomposition},
the $L$-steps ahead prediction of $\vct{x}_d$ is described as follows:
\begin{align}
\vct{x}_d(t_0 + L\Delta t) &= \ee^{{\sf A}_d L\Delta t} \vct{x}_d(t_0)\nonumber\\
&= {\sf V}_d
\left[
  \begin{array}{ccc}
    \ee^{\lambda_1 L\Delta t} &&\\
    &\ddots&\\
    &&\ee^{\lambda_d L\Delta t}
\end{array}
\right]
{\sf V}_d^{-1}\vct{x}_d(t_0)\nonumber\\
&= \sum_{i=1}^{d} a_{i, d}(t_0) \ee^{\lambda_i L\Delta t} \vct{V}_i
\label{eq:seq_evol}
\end{align}
where $a_{i, d}(t_0)$ is the $i$-th component of ${\sf V}_d^{-1}\vct{x}_d(t_0) \in \mathbb{C}^d$.
\eqref{eq:seq_evol} implies that each subsequence evolves with the initial time-dependent coefficients $a_{i,d}(t_0)$ on the space spanned by $\{\vct{V}_i\}$.
Here, we assume that the underlying continuous-time signal $y(t)$ can be represented as a finite linear combination of exponential modes $y(t) = \sum_j \exp(\lambda_j^{\mathrm{true}} t)$,
which corresponds to a system with a discrete spectrum $\{\lambda_j^{\mathrm{true}}\}$.
This assumption is commonly employed in finite-dimensional linear approximations of nonlinear dynamical systems \cite{pan2020structure}.
Then, each $\vct{V}_i = [v_{i,1}, \ldots, v_{i,d}]$ needs to be a sequence of discrete-time exponential functions:
\begin{align}
  \vct{V}_i \simeq b_i\left[
  \begin{array}{c}
  \exp\{\nu_i((d-1)\Delta t)\}\\
  \exp\{\nu_i((d-2)\Delta t)\}\\
  \vdots\\
  1
  \end{array}
  \right],\,\,\, b_i, \nu_i\in\mathbb{C}.
  \label{eq:seq_exp}
\end{align}
In contrast to $\exp(\lambda_i t)$, which denotes the map from one subsequence to another,
the component $\exp(\nu_i t)$ measures {\em transient} responses contained in each subsequence.
Thus, the real part of $\exp(\nu_i t)$ corresponds to the decaying rate of the responses.
We propose the perspective that the decaying rate indicates the lengths (or timescales) of temporal patterns. 
By combining \eqref{eq:seq_evol} and (\ref{eq:seq_exp}) and by setting $L = 0$,
we approximate original $y(t)$ by superposing the components $\exp(\nu_i t)$\footnote{While we present the formulation for scalar outputs for clarity, the scheme naturally extends to vector-valued outputs by applying the same construction element-wise.}: for $n = 0, 1, \ldots, d-1$,
\begin{align}
  y(t - n\Delta t) \simeq&
  \sum_{i=1}^{d} a_{i,d}(t) \exp \left(
    \nu_i (d - n - 1)\Delta t
    \right)\nonumber\\
    =& \sum_{i=1}^{d} \tilde{a}_{i,d}(t)\exp
      \left(
      - \nu_i n\Delta t
      \right),
  \label{eq:delay_expansion}
\end{align}
where $t_0$ is rewritten by $t$,
and the coefficient $\tilde{a}_{i,d} (t) := a_{i,d}(t) \exp \{\nu_i (d-1)\Delta t\}$ indicates how the component with $\nu_i$ is dominant in the interval $[t-d\Delta t, t]$.
Assuming each $\nu_i$ is distinct so that $\{\exp(\nu_i t)\}$ are linearly independent functions on $[t-d\Delta t, t]$,
we propose to estimate $\tilde{a}_{i,d}(t)$ by projecting $y(t)$ onto $\exp(\nu_i t)$:
\begin{align}
  \tilde{a}_{i,d}(t) \simeq \sum_{n=0}^{d-1} y(t - n\Delta t)
  \exp(\nu_i n \Delta t)
  =: \tilde{A}_{i,d}(t)
  \label{eq:fit}.
\end{align}
The $\tilde{A}_{i,d}(t)$ helps to find dominant transient responses on $[t - d\Delta t, t]$.
Below, by representing $\nu_i = \sigma_i + \ii \omega_i$
($\sigma_i\in\mathbb{R}$ and $\omega_i\in[0, 2\pi]$),
we introduce the $i$-th {\em time constant}
$T_i := 1/\sigma_i$ as the timescale of the pattern that appears on $[t_0 - d\Delta t, t_0]$ and involves $\tilde{A}_{i,d}(t)$.

\subsection{Implementation}
\label{sec:algorithm}

We have developed an algorithm to characterize lengths of temporal patterns based on the idea mentioned above.
The challenge now is to determine an appropriate value for $d$,
which represents the degree of the delay coordinates.
In terms of extracting temporal patterns, $d$ should be greater than the length of the longest pattern of interest.
However, a large value of $d$ should be avoided as it can lead to overfitting of ${\sf A}_d$,
which can negatively impact the extraction of essential dynamic properties.
To address this challenge, we suggest using the Akaike Information Criterion (AIC; see Refs.~\cite{Akaike1973,Hastie}) to determine the appropriate value of $d$.

Let us prepare mathematical formulations for identifying ${\sf A}_d$ and determining $d$.
Consider the discrete-time data $y[n]$ ($n=0, 1, \ldots, N-1$) sampled from the continuous time series $y(t)$ and its delay coordinate vector $\vct{x}_d[n] = [y[n], y[n-1], \ldots, y[n-d+1]]^\top$.
We introduce the data matrices ${\sf X}_d^{(M_1,M_2)}, {\sf Y}_d^{(M_1,M_2)}\in \mathbb{R}^{d \times (M_2-M_1)}$, denoted by
\begin{align}
  {\sf X}_d^{(M_1,M_2)} &= \left[
    \vct{x}_d[M_1-1],\ldots, \vct{x}_d[M_2-2]
    \right]\label{eq:exp_matrix},\\
  {\sf Y}_d^{(M_1, M_2)} &= \left[
    \vct{x}_d[M_1],\ldots, \vct{x}_d[M_2-1]
    \right]\label{eq:obj_matrix},
\end{align}
where each sample $\vct{x}_d[n]$ is stacked as a column vector, 
so that ${\sf X}_d^{(M_1,M_2)}$ and ${\sf Y}_d^{(M_1,M_2)}$ are matrices of size (state dimension $\times$ number of samples).
Given a certain degree $d$, ${\sf A}_d$ is derived as a temporally discretized form $\tilde{\sf A}_d$ by minimizing the following 1-ahead prediction error:
\begin{align}
  E_d = \left\|
    {\sf Y}_d^{(d,N)} - \tilde{\sf A}_d {\sf X}_d^{(d,N)}
    \right\|_\mathrm{F}^2
  \label{eq:reg_error},
\end{align}
where $\|\cdot\|_\mathrm{F}$ denotes the Frobenius norm.
The minimization of the cost function (\ref{eq:reg_error}) with respect to  $\tilde{\sf A}_d$ reduces to a standard least-squares problem. 
Specifically, solving
\begin{align}
  \tilde{\sf A}_d \, 
  \left( {\sf X}_d^{(d,N)} ({\sf X}_d^{(d,N)})^\top \right)
  &= {\sf Y}_d^{(d,N)} ({\sf X}_d^{(d,N)})^\top,
  \label{eq:normal_eq}
\end{align}
yields the closed-form solution
\begin{align}
  \tilde{\sf A}_d 
  &= {\sf Y}_d^{(d,N)} ({\sf X}_d^{(d,N)})^\top 
     \left( {\sf X}_d^{(d,N)} ({\sf X}_d^{(d,N)})^\top \right)^{-1}.
  \label{eq:ls_solution}
\end{align}
This corresponds to the ordinary least-squares estimator\footnote{In the present study, no additional regularization (e.g., L1 or L2 penalty) was imposed, 
since the datasets considered here allowed for a stable estimation. 
Nevertheless, regularization could be incorporated into \eqref{eq:reg_error} in future extensions to improve robustness when ${\sf X}_d^{(d,N)}$ is ill-conditioned.}.
However, it is necessary to regularize \eqref{eq:reg_error} to avoid the overfitting of $\tilde{\sf A}_d$ to the training dataset $\{{\sf X}_d, {\sf Y}_d\}$.
We use the prediction horizon $L$ larger than 1 to generate a test dataset different from $\{{\sf X}_d, {\sf Y}_d\}$
and introduce the AIC, which is the relative quality of statistical models for a given dataset.
Under a fixed $L > 1$, the AIC for the $L$-ahead linear regression model is given by
\begin{align}
  \mathrm{AIC}(d) =& 
  \frac{\left\|
  {\sf Y}_d^{(d+L-1,N)} - {\sf C}_d \tilde{\sf A}_d^L {\sf X}_d^{(d,N-L+1)}
  \right\|_\mathrm{F}^2
  }{N-d-L}
  \nonumber\\
  &+ \frac{2d}{N-d-L}\sigma^2
  \label{eq:AIC},
\end{align}
where ${\sf C}_d = [\overbrace{1, 0, \ldots, 0}^{d}]$ and $\sigma^2$ represents the estimate of noise variance.
The $\sigma^2$ is estimated by signal analysis or tuned as a regularization parameter.
The design parameter $L$ represents the prediction horizon in the information criterion.
In principle, there is no universally optimal choice of $L$.
However, its selection significantly affects the behavior of (\ref{eq:AIC}).
A too small $L$ (e.g., $L=1$) trivializes the criterion because the prediction horizon is too short, whereas too large $L$ (e.g., $L\geq 10$) tends to destabilize the evaluation due to error accumulation in long-term prediction.
In our experiments, we empirically observed that $L=5$ provides a reasonable trade-off
between stability, predictive capability, and the convexity of the information criterion.
Therefore, throughout this paper, we fixed $L=5$ for all experiments.
A comparison of different values of $L$ (e.g., $L=1, 5, 10$) is provided in Appendix~B to illustrate this trade-off.

Based on the formulations, we develop the Algorithm~1 to characterize timescales of temporal patterns inside subsequences.
\begin{algorithm}[tb]
\caption{Estimation of temporal patterns with optimal delay order $d^*$}
\label{alg:est_pattern}
\KwIn{Discrete-time samples $y[n]$, candidate set $\Omega$ of $d$, prediction horizon $L>1$}
\KwOut{Eigenvalues $\{\tilde{\nu}_i\}$, amplitudes $\{\tilde{A}_{i,d^*}[n]\}$, and time constants $\{\tilde{T}_i\}$}

\For{$d \in \Omega$}{
  Identify $\tilde{{\sf A}}_d$ by solving the least-squares problem in \eqref{eq:reg_error}\;
  Compute $\mathrm{AIC}(d)$ using \eqref{eq:AIC}\;
}
Select $d^* = \arg\min_{d \in \Omega} \mathrm{AIC}(d)$\;
Obtain eigenvectors $\{\tilde{\vct{V}}_i\}_{i=1,\ldots,d^*}$ of $\tilde{{\sf A}}_{d^*}$\;
\For{$i=1$ \KwTo $d^*$}{
  Fit a discrete-time exponential curve $\big[\exp(\tilde{\nu}_i(d-1)), \exp(\tilde{\nu}_i(d-2)), \ldots, 1\big]$ to $\tilde{\vct{V}}_i$ to estimate $\tilde{\nu}_i$\;
  Compute time constant $\tilde{T}_i = 1 / \mathrm{Re}[\tilde{\nu}_i]$\;
  \For{each time index $n$}{
    Calculate amplitude $\tilde{A}_{i,d^*}[n] = \vct{x}_{d^*}[n]^\top\tilde{\vct{V}}_i$\;
  }
}
\end{algorithm}
In practice, the dominant peaks were initially identified heuristically,
by balancing peak height and spectral isolation from continuous components. 
This procedure was sufficient to highlight the main temporal scales in the experiments below,
but it inherently involves a degree of subjectivity.
To improve reproducibility and clarity, we further provide a formalized selection algorithm (Algorithm~2 in Appendix~C) that reproduces the same peak candidates in a systematic manner.

Finally, we note how the DEF quantities are used downstream:
the triplets $\{(\tilde{\nu}_i,\tilde{\vct{V}}_i,\tilde{A}_{i,d^*}[n])\}$ act as \emph{pattern-level descriptors} that bridge raw signals and subsequent analysis/design.
They summarize dominant temporal scales per subsequence, enable search/indexing across runs, and provide compact targets for validating or refining physics-based models and controller settings.

\section{Numerical verification with toy model}
\label{sec:numerical}

\begin{figure}[t]
  \begin{center}
    \includegraphics[width=0.95\hsize]{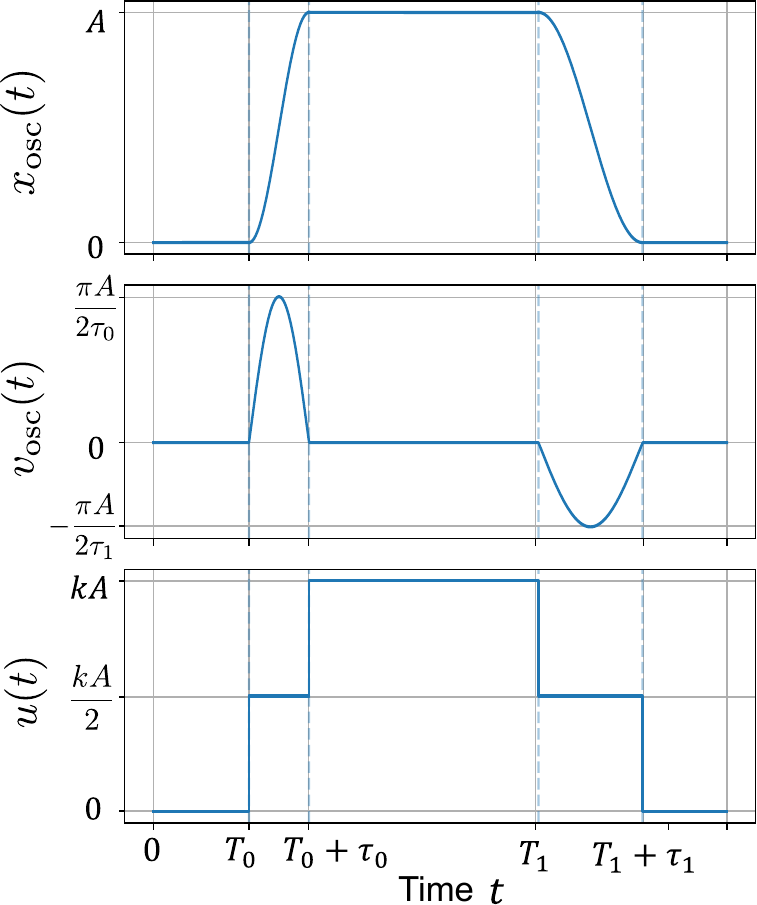}
    \caption{Trajectory of $x(t)$ and $v(t)$ in \eqref{eq:toy_model} driven by the PWC input $u(t)$.}
    \label{fig:toy_model}
  \end{center}
\end{figure}
In this section, we verify that the proposed method identifies two temporal patterns with their different lengths in numerical data derived from a toy model.
As a fundamental model of industrial mechanical systems,
we consider a simple harmonic oscillator driven by a piecewise constant (PWC) force $u(t)$ and a switching mass $m(t)$:
\begin{align}
  \frac{\dd}{\dd t} \left[
    \begin{array}{c}
      x\sub{osc}(t)\\
      v\sub{osc}(t)
    \end{array}
    \right] = \left[
    \begin{array}{cc}
      0 & 1 \\
      -\frac{k}{m(t)} & 0
    \end{array}
    \right]
    \left[
    \begin{array}{c}
      x\sub{osc}(t)\\
      v\sub{osc}(t)
    \end{array}
    \right] + \left[
      \begin{array}{c}
        0\\
        \frac{u(t)}{m(t)}
      \end{array}
      \right],
  \label{eq:toy_model}
 \end{align}
where $x\sub{osc}(t)$ is the oscillator's position,
$v\sub{osc}(t)$ is the velocity, $k$ is the spring constant.
We set the initial condition $x\sub{osc}(0) = v\sub{osc}(0) = 0$.

Next, we model the mass $m(t)$ as a PWC function switching 
between $m_0$ and $m_1$. 
This switching governs the oscillation timescales: 
$\tau_0 := \pi\sqrt{m_0/k}$ and $\tau_1 := \pi\sqrt{m_1/k}$ denote the half periods governing the rise and fall segments, respectively. 
By allowing $m(t)$ to change, the system exhibits trajectories with multiple 
characteristic frequencies.

On this basis, we introduce the external force $u(t)$ as a \emph{designed} PWC input.
Specifically, $u(t)$ switches through five levels $0 \!\to\! \tfrac{kA}{2} \!\to\! kA \!\to\! \tfrac{kA}{2} \!\to\! 0$ so that the rise and fall segments each span one half natural period of the oscillator, yielding a half-sine rise to $x\sub{osc}\!=\!A$, a flat plateau at $A$, and a half-sine return to $0$. The schedule is
\begin{equation}
u(t)=
\begin{cases}
0, & t < T_0,\\[4pt]
\dfrac{kA}{2}, & T_0 \le t < T_0+\tau_0,\\[6pt]
kA, & T_0+\tau_1 \le t < T_1,\\[6pt]
\dfrac{kA}{2}, & T_1 \le t < T_1 + \tau_1,\\[6pt]
0, & t \ge T_1 + \tau_1,
\end{cases}
\label{eq:designed_pwc_input}
\end{equation}
where $\tau_0 := \pi\sqrt{m_0/k}$ and $\tau_1 := \pi\sqrt{m_1/k}$ denote the half periods for the rise and fall segments (using the corresponding mass values).
This PWC excitation $u(t)$ displaces the oscillator from the trivial equilibrium, producing piecewise sinusoidal responses: the trajectory of $x(t)$ consists of a half-sine rise to $A$, a constant plateau at $A$, and a half-sine decay to $0$.
Thus, this section considers a two-input system, $(u(t),\,m(t))$.
\begin{figure}[t]
  \begin{center}
    \includegraphics[width=0.8\hsize]{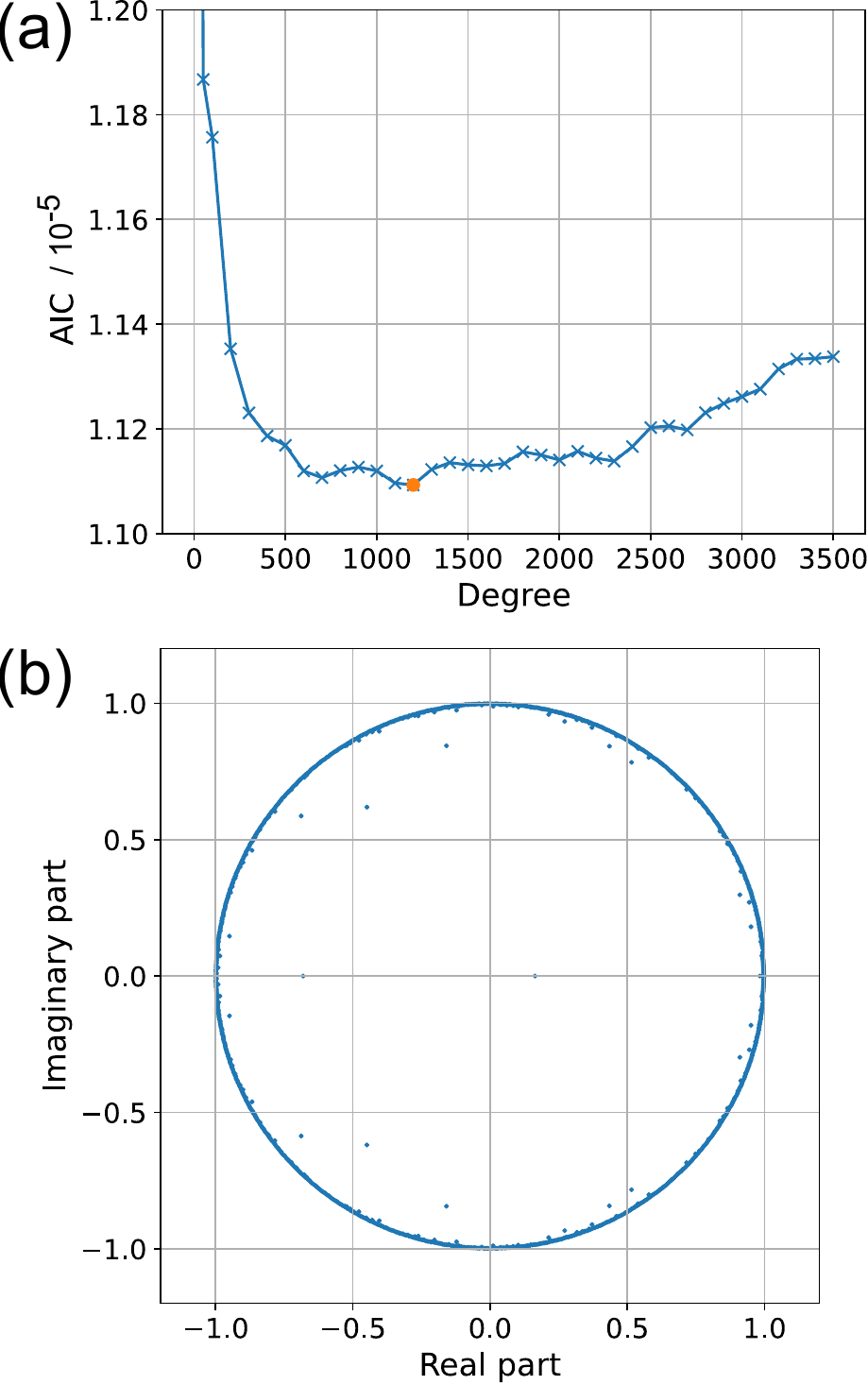}
    \caption{Preliminary results for the toy model data:
    (a) Degree $d$ vs. $\mathrm{AIC}(d)$, where the orange circle denotes the optimal value $d^*$, and
    (b) Eigenvalues of $\tilde{\sf A}_d$ with $d = 1200$.}
    \label{fig:preliminary_toy_model}
  \end{center}
\end{figure}
Figure~\ref{fig:toy_model} illustrates $x\sub{osc}(t)$ and $v\sub{osc}(t)$.
Below, we treat $x\sub{osc}(t)$ as the target time series $y(t)$ analyzed by DEF.

The method for verifying that the DEF can characterize temporal patterns is delineated below.
We set $A = 1$, $\tau_0 = 1000$, $\tau_1 = 100$, $N = 15000$, $T_0 = 5000$, and $T_1 = 10000$ and generate a dataset $\{y[0], \ldots, y[N-1]\}$ with the sampling period $\Delta t = 1$.
Also, we added the Gaussian noise with its variance of $10^{-6}$ to the generated data.
We applied DEF to the data to compute the time constant $\tilde{T}_i$ and the associated $\tilde{A}_{i,d}[n]$,
where $\sigma^2$ in \eqref{eq:AIC} is set at $5\times 10^{-6}$ as it is of the same order as the variance of the noise added above.
The fitting onto $\tilde{\vct{V}}_i$ was conducted by \texttt{scipy.signal.find\_peaks} in Python.
Then, we extract the time constant $\tilde{T}_i$ associated with large $\tilde{A}_{i,d}[n]$ and investigate its validity in comparison with the $\tau_0$ or $\tau_1$.
Note that our method does not impose a strict requirement on the sampling period $\Delta t$ or the number of samples $N$.
However, as in general signal processing, if $\Delta t$ is chosen too large relative to the characteristic timescales of the dynamics, the temporal patterns cannot be resolved and extraction becomes difficult.
In practice, it suffices to set $\Delta t$ small enough to resolve these timescales.
In our experiments, we selected $\Delta t = 1$ and $\tau_0 = 1000$ (or $\tau_1 = 100$), which were adequate to capture both oscillatory and decay behaviors.

\begin{figure*}[tb]
  \begin{center}
    \includegraphics[width=0.9\hsize]{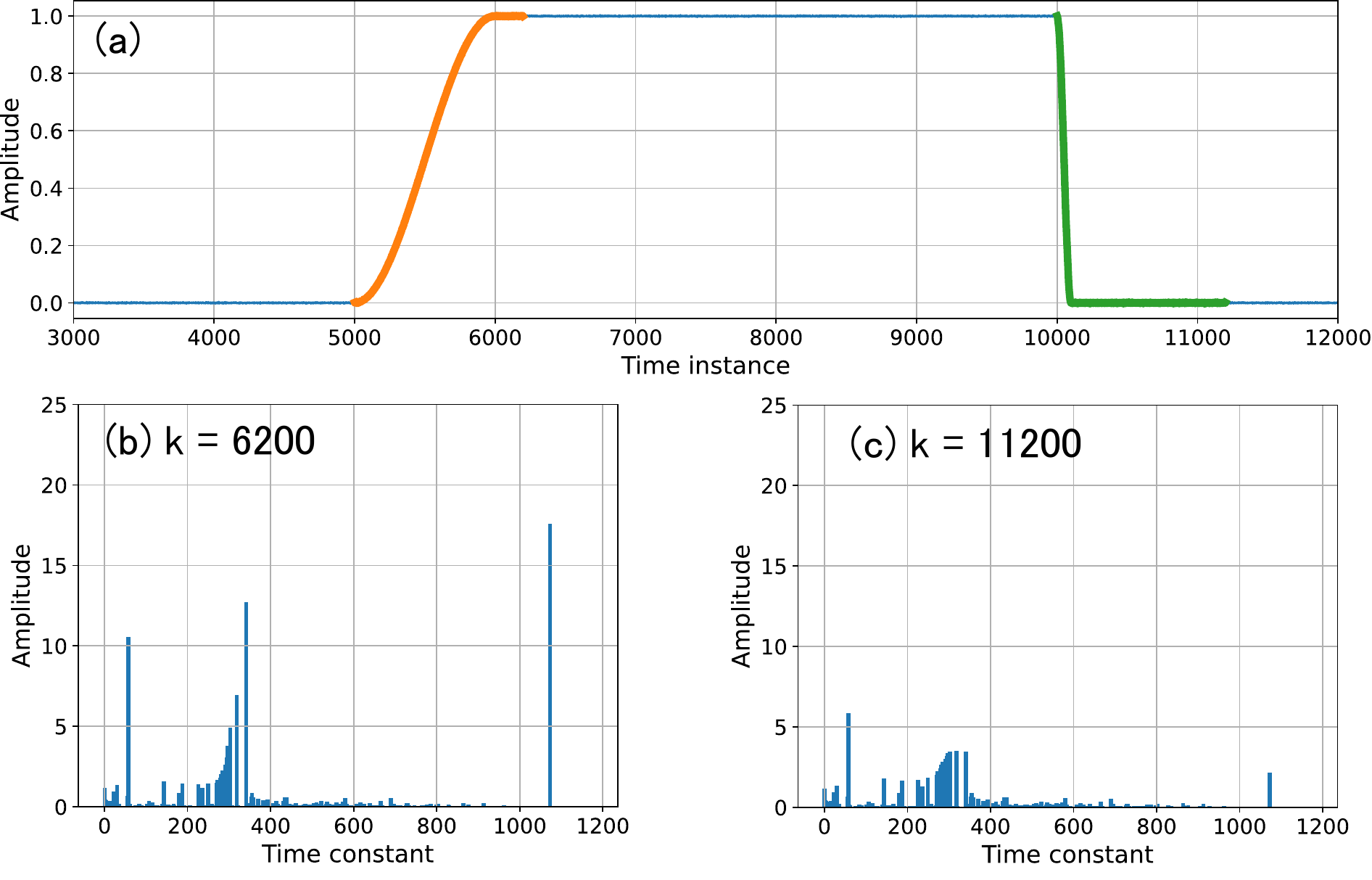}
    \caption{Main results for the toy model data:
    (a) The target subsequences $\vct{x}_{1200}[6200]$ and $\vct{x}_{1200}[11200]$ highlighted by the bold orange and green lines, respectively.
    (b,c) The magnitude $\tilde{A}_{i,d}[n]$ at each time constant $\tilde{T}_i$ for (b) $n = 6200$ and (c) $n = 11200$.}
    \label{fig:toy_model_result}
  \end{center}
\end{figure*}
Let us show the preliminary results.
Figure~\ref{fig:preliminary_toy_model}~(a) shows the AIC values for each degree $d$, which were set at $\{100, 200, ..., 2000\}$.
As $d$ increased, the AIC values decreased and reached their minimum at $d = 1200$.
This value of $d$ is close to the length of the first sinusoidal curve ($\tau_0 = 1000$), which is the longest temporal pattern in the data. Therefore, we adopted $d = 1200$ to characterize the temporal patterns in the rest of this section.
Also, \figref{fig:preliminary_toy_model}~(b) shows the eigenvalues $\{\tilde{\lambda}_i\}$ of $\tilde{\sf A}_d$ computed with $d = 1200$.
The $\tilde{\lambda}_i$ lie close to the unit circle, where the distance between any pair of eigenvalues took its minimum of $3.52\times 10^{-4}$.
This implies that all the $\tilde{\lambda}_i$ are distinct so that $\tilde{\sf A}_d$ is factorized with $d$ linearly independent eigenvectors, which is the assumption we hold above.

Then, we show the main results of DEF.
We focused on the two subsequences $\vct{x}_{1200}[6200]$ and $\vct{x}_{1200}[11200]$ that contained the sinusoidal curves highlighted by bold orange and green lines in \figref{fig:toy_model_result}~(a).
The panels~(b) and (c) show the magnitude $\tilde{A}_{i,d}[n]$ at each time constant $\tilde{T}_i$ for $n= 6200$ and $n = 11200$, respectively.
The graph in panel (b) indicates that the amplitude reaches its highest peak at $\tilde{T}_i = 1072.9418$,
which suggests that DEF identifies the first sinusoidal curve with its length $\tau_0 = 1000$.
Similarly, panel (c) shows that the amplitude takes the highest peak at $\tilde{T}_i = 75.3143$,
which is close to the length $\tau_1 = 100$ of the second sinusoidal curve.
Here, in both panels (b,c),
the amplitude takes large values around $\tilde{T}_i = 341.0002$,
but no clear peak appears around this time constant.
This finding implies that the exponential components with time constants close to $\tilde{T}_i = 341.0002$ reflect residual content not associated with a single dominant timescale.

Note that the estimation accuracy depends on how the delay order $d$ is aligned with the time scale of the underlying temporal pattern. 
In the above results with $d = 1200$, the estimation error for the long-scale pattern ($\tilde{T}_i = 1072.94$ vs. $\tau_0 = 1000$) is relatively smaller than that for the short-scale pattern ($\tilde{T}_i = 75.31$ vs. $\tau_1 = 100$). 
This indicates that shorter windows $d$ are preferable for short-scale patterns, while longer windows are more suitable for long-scale patterns. 
In practice, one can first apply the method with a broad range $[d\sub{min}, d\sub{max}]$ to capture the main peaks, and then re-run the estimation with a restricted window length range around the identified scale to refine accuracy.
This two-step strategy was found to be sufficient for engineering use cases.

From the above results and discussion,
DEF can quantify how long sinusoidal curves are dominant at each time instance.

\section{Demonstration with industrial data}
\label{sec:demo}

This section provides a demonstration of how DEF can extract multiple patterns of different lengths from complex time series data obtained from a manufacturing machine. 

\begin{figure}[t]
   \begin{center}
    \includegraphics[width=0.8\hsize]{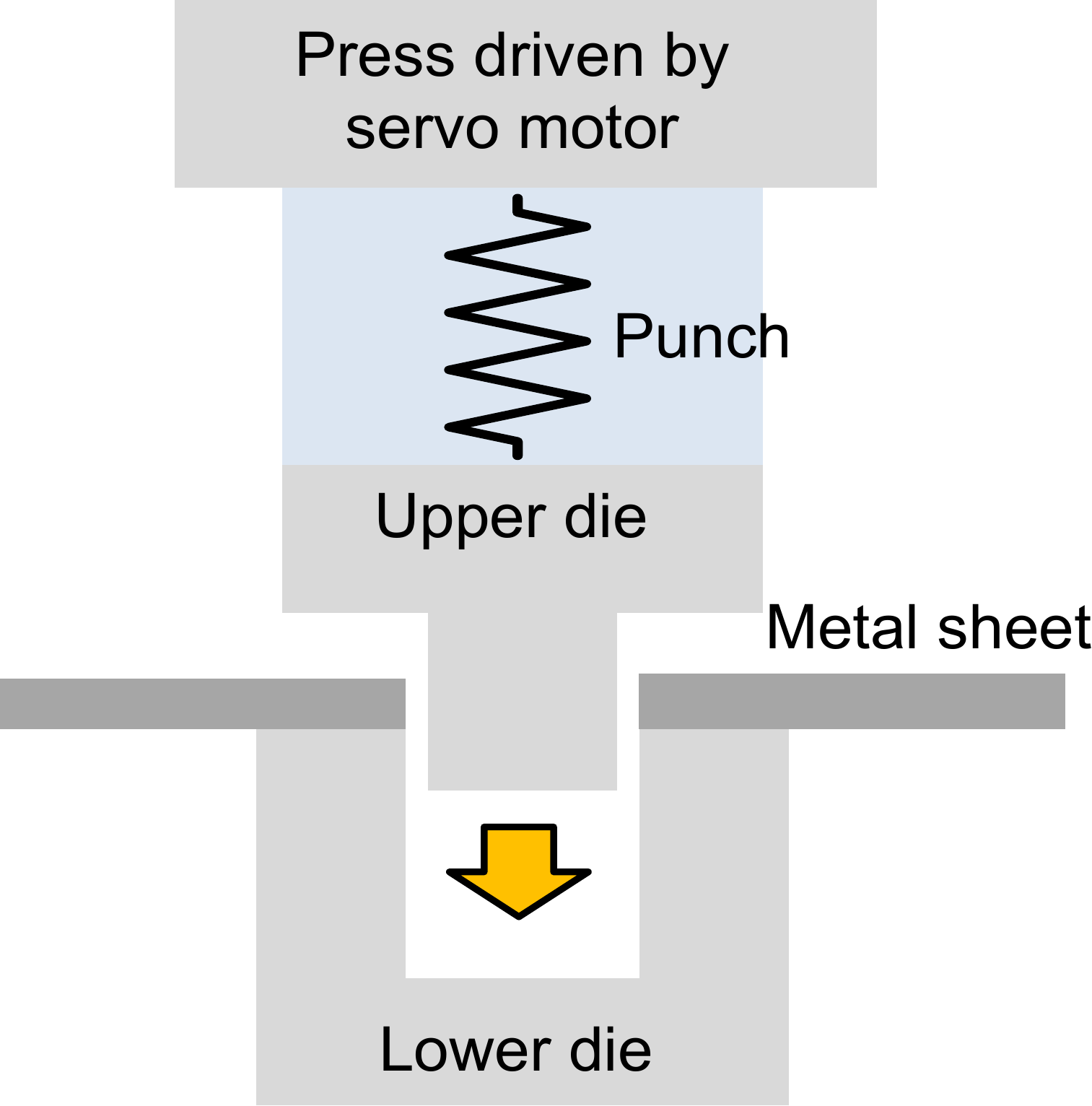}
     \caption{Schematic diagram of turret punch blanking a metal sheet.}
     \label{fig:punch}
   \end{center}
 \end{figure}

The demonstration focuses on a turret punch, which is a type of punch press used for metal forming. 
Figure~\ref{fig:punch} shows a schematic diagram of the target turret punch.
\begin{figure*}[!t]
   \begin{center}
     \includegraphics[width=\hsize]{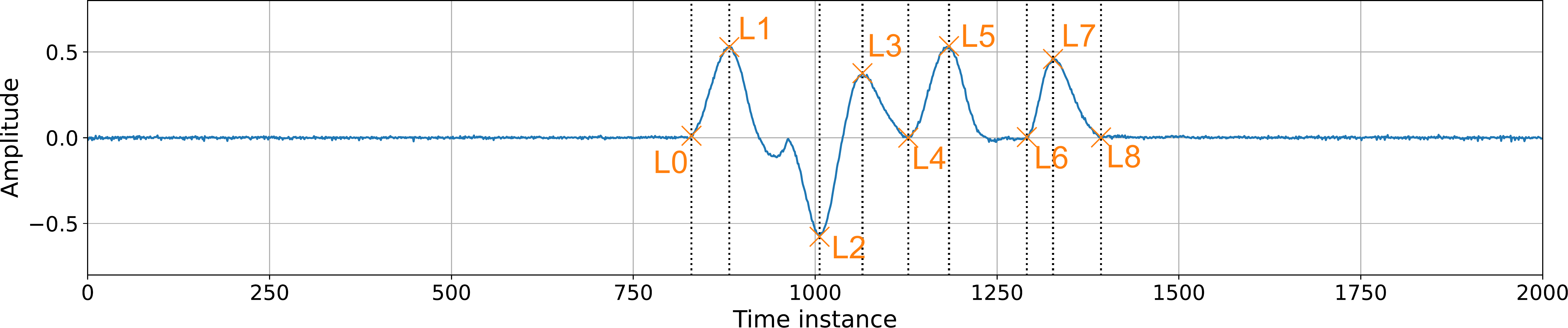}
     \caption{Target current data measured in the servo motor that drives the press, where the representative locations are denoted by L0, L1, \ldots, L8.}
     \label{fig:punch_data}
   \end{center}
 \end{figure*}
 The press is driven by a servo motor and punches an upper die to pierce sheet metal.
We collected data on the electric current of the servo motor while driving the press.
Figure~\ref{fig:punch_data} shows the current data, with 2000 data points, where a metal sheet was punched at $n= 830$.
During the punching process, the reference signal for controlling the vertical position of the upper die jumped to accelerate it, causing intermittent current fluctuations as shown in the figure.
Using the reference signal as the input $\vct{u}(t)$ in \eqref{eq:dyn}, the current data can be naturally described as $y(t)$ generated by \eqref{eq:dyn} and \eqref{eq:obs}.
As shown in the figure, the representative locations of the data are denoted by L0, L1, \ldots, L8 to characterize the fluctuation.
We verify the possibility of DEF to identify temporal patterns determined by these locations of the data.
Note that the setting of DEF was as in \secref{sec:numerical};
$\sigma^2$ was set at $10^{-3}$ in the same order as the measurement noise,
and the fitting onto $\tilde{\vct{V}}_i$ was conducted by \texttt{scipy.signal.find\_peaks}.

\begin{figure}[t]
  \begin{center}
    \includegraphics[width=0.8\hsize]{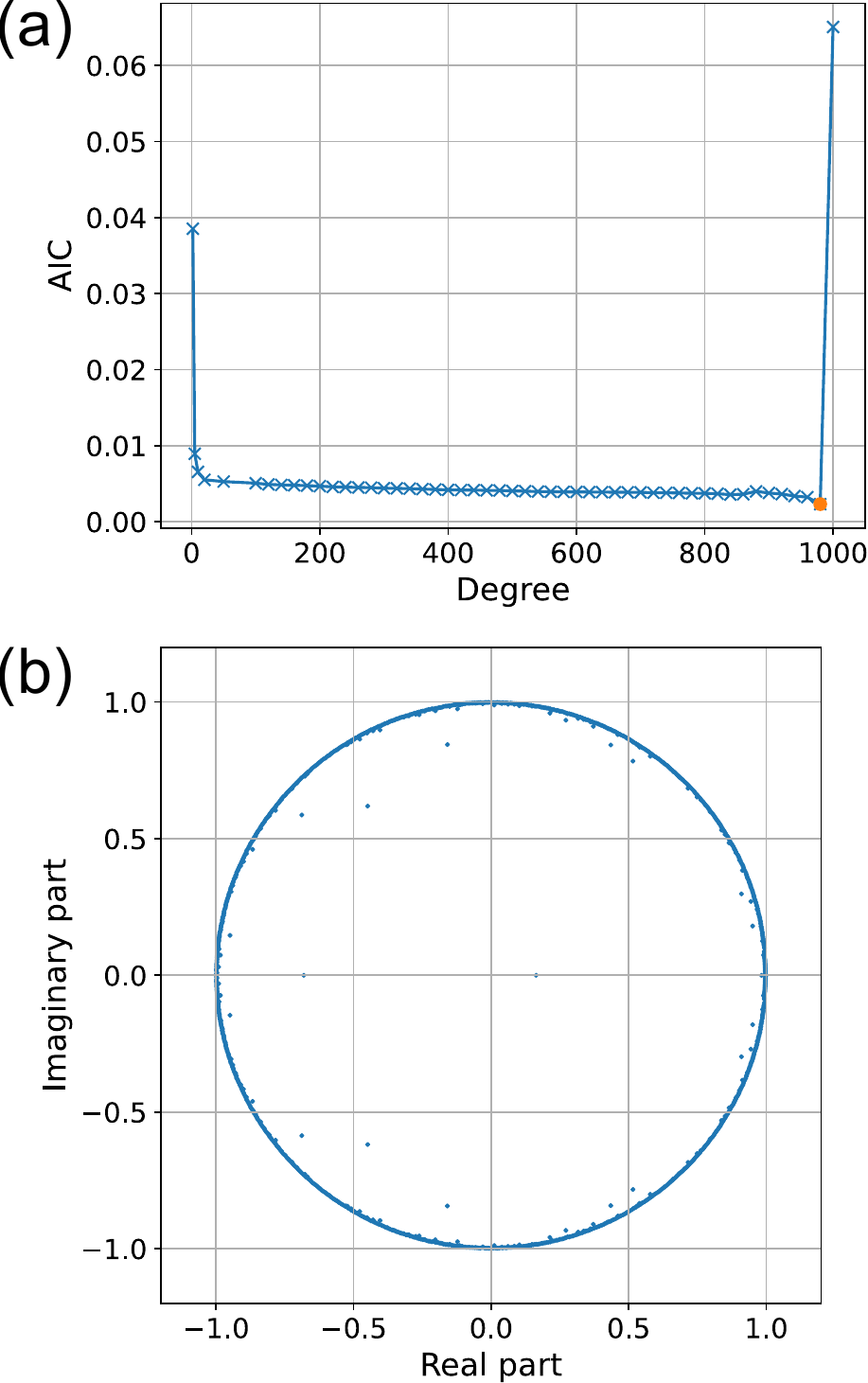}
    \caption{Preliminary results for the turret punch data:
    (a) The degree $d$ vs. $\mathrm{AIC}(d)$, where the orange circle denotes the optimal value $d^*$, and
    (b) Eigenvalues of $\tilde{\sf A}_d$ with $d = 980$.}
    \label{fig:preliminary_press}
  \end{center}
\end{figure}
Let us show the preliminary results for DEF.
Figure~\ref{fig:preliminary_press}(a) shows the AIC for each $d$.
The AIC took its minimum at $d = 980$,
corresponding to the duration while the fluctuation occurred.
Therefore, we adopted $d = 980$ to conduct the DEF analysis.
Also, \ref{fig:preliminary_press}(b) shows the eigenvalues $\tilde{\sf A}_d$ for $d = 980$,
where the eigenvalues are uniformly distributed around the unit circle
and the distance between any pair of eigenvalues took its minimum $1.35972\times 10^{-3}$.
Thus, as in the \secref{sec:numerical},
$\tilde{\sf A}_d$ is factorized with $d$ linearly independent eigenvectors.

\begin{figure}[!t]
  \begin{center}
    \includegraphics[width=0.9\hsize]{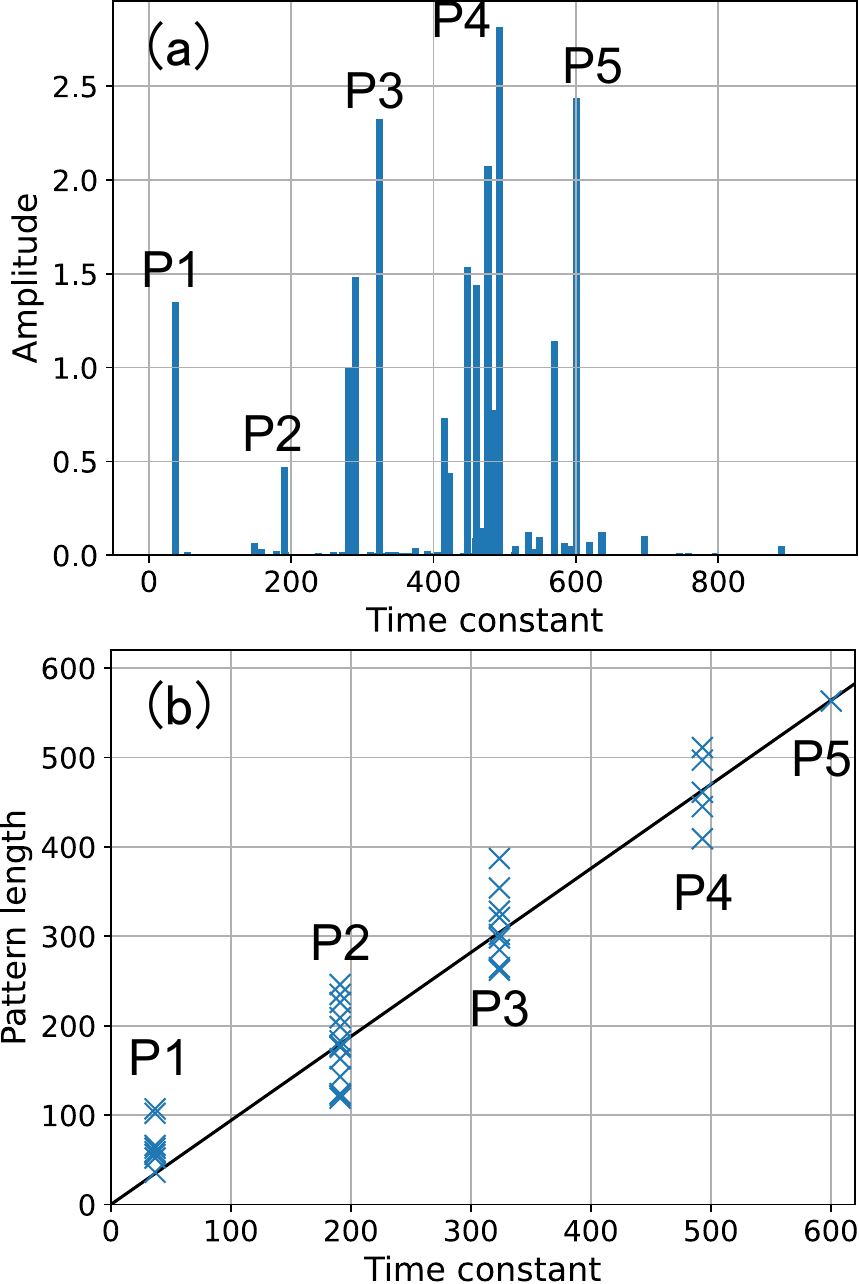}
    \caption{DEF results for the turret punch:
    (a) magnitude $\tilde{A}_{i,d}[n]$ for each time constant $T_i$ at $n= 1800$ and
    (b) pattern lengths $L_{i,j}$ ($i, j \in \{0, \ldots, 8\}$) computed from the turret punch data,
    which are separately plotted for their nearest time constant that corresponds to the peaks P1 to P5.}
    \label{fig:press_result}
  \end{center}
\end{figure}

Figure~\ref{fig:press_result}~(a) shows the magnitude $\tilde{A}_{i,d}[n]$ at each time
constant $\tilde{T}_i$ ($i = 1, 2, \ldots, d$) for $n= 1800$.
The distribution of the $\tilde{A}_{i,d}[n]$ has a finite number of peaks,
implying that DEF can extract dominant bases from the $d$ ($=980$) original bases.
We address the six peaks called P1--P5 in the panel (a),
whose time constants are 36.85, 191.1, 323.8, 492.9, and 600.3, respectively.
The time constant for each peak is close to the length of different temporal patterns included in the current data;
e.g., the time constant 600.3 of P5 is almost equal to the distance between L0 and L8.
To clearly show this,
we compute pattern lengths $L_{i,j}$ directly from the real turret punch dataset, defined as the empirical distance between location L$i$ and L$j$
(e.g., $L_{0,1}$ denotes the distance between L0 and L1).
We displayed them separately for their nearest time constant in \figref{fig:press_result}~(b).
The pattern length can be approximated as a linear function of the time constant with its slope of 0.98,
implying that the derived time constant represents the length of patterns inherent in the turret punch data.
Note that the approximation line deviates from the pattern lengths for P1.
Recalling $T_i = 36.85$ for P1 is of a smaller order than $d = 980$,
the deviation implies that DEF is suitable to the identification of an overall (long-term) trend in a subsequence compared to a short-term one.

From the above results, we contend that DEF can characterize multiple patterns with different lengths.

\section{Comparative analysis}
\label{sec:ssa_dmd}

In order to evaluate the effectiveness of the proposed DEF method, we compare its performance with two widely used decomposition techniques: Singular Spectrum Analysis (SSA) and delay-embedded Dynamic Mode Decomposition (Delay-DMD).
For each method, both the toy model data in \secref{sec:numerical} and the turret punch data in \secref{sec:demo} are analyzed.

Note that conventional time-series data mining approaches such as motifs \cite{Lin2002,Lam2011}, discords \cite{Keogh2005,Mueen2010}, and shapelets \cite{Ye2009,Yamaguchi2020} essentially rely on repetition or singularity.
These criteria are ill-suited to the present problem setting, where only a limited number of one-shot pulsive patterns are available.
In contrast, SSA \cite{Hassani2007,Golyandina2001} and DMD \cite{LeClainche2017,Brunton2017,Schmid2010} share a common reliance on Hankel (trajectory) matrices constructed directly from the observed signal, and thus do not require multiple instances of the same pattern.
Accordingly, we adopt SSA and a delay-embedded variant of DMD as \emph{Hankel-based} baselines and compare the time scales they extract against those of DEF.
For fairness, the embedding dimension in SSA and Delay-DMD is fixed equal to the DEF delay order $d^\ast$ (note that $d^\ast$ differs between the toy system and the turret dataset).

\vspace{0.5em}
\subsection{Baselines via Hankel-structured methods}
\label{sec:baseline}

\subsubsection{Singular Spectrum Analysis (SSA)}
\label{subsec:ssa}

Let $y[n]$ ($n=0,1,\dots,N-1$) be the discrete-time signal sampled from $y(t)$ with sampling period $\Delta t$.
Given a window length $d$ (here set to $d=d^\ast$ to match DEF), we construct the Hankel (trajectory) matrix
\begin{equation}
  \mathbf{X}_{\mathrm{SSA}}
  =
  \begin{bmatrix}
    y[0]   & y[1]   & \cdots & y[K-1] \\
    y[1]   & y[2]   & \cdots & y[K]   \\
    \vdots & \vdots & \ddots & \vdots \\
    y[d-1] & y[d]   & \cdots & y[N-1]
  \end{bmatrix},
  \label{eq:ssa_hankel}
\end{equation}
where $K := N-d+1$.
Applying the SVD to $\mathbf{X}_{\mathrm{SSA}}$,
\begin{equation}
  \mathbf{X}_{\mathrm{SSA}}
  \;=\;
  \sum_{i=1}^{d} \sigma^{\mathrm{SSA}}_{i}\,\mathbf{u}^{\mathrm{SSA}}_{i}\,(\mathbf{v}^{\mathrm{SSA}}_{i})^{\!\top},
  \label{eq:ssa_svd}
\end{equation}
yields singular triplets $\big(\sigma^{\mathrm{SSA}}_{i},\mathbf{u}^{\mathrm{SSA}}_{i},\mathbf{v}^{\mathrm{SSA}}_{i}\big)$.
Following standard SSA, \emph{reconstructed components} (RCs) are obtained by diagonal averaging of selected outer products
$\sigma^{\mathrm{SSA}}_{i}\,\mathbf{u}^{\mathrm{SSA}}_{i}\,(\mathbf{v}^{\mathrm{SSA}}_{i})^{\!\top}$.
We denote the resulting one-dimensional RC signals by $\{ \mathrm{RC}_{i}[n] \}_{i=1}^{d}$.
Since SSA has no unified, principled method to identify the \emph{time scale} of each $\mathrm{RC}_{i}[n]$, we qualitatively test the applicability of SSA to identify pattern length in the toy model data and the turret punch data by showing $\mathrm{RC}_{1}[n]$ and $\mathrm{RC}_{2}[n]$ as representative waveforms.

\vspace{0.5em}
\subsubsection{Delay-embedded DMD (Delay-DMD)}
\label{subsec:delay_dmd}

We apply DMD to the delay-embedded snapshots built from $y[n]$ with the same embedding length $d = d^\ast$ as DEF.
Following the standard (projected) DMD pipeline, we compute a rank-$r$ truncated SVD of ${\sf X}_d^{(d,N)}$,
\begin{align}
  {\sf X}_d^{(d,N)}
  \;\approx\;
  {\sf U}_r^{\mathrm{DMD}} \,\Sigma_r\, ({\sf V}_r^{\mathrm{DMD}})^{\!\top},
\end{align}
and estimate the reduced operator
\begin{align}
  {\sf A}^{\mathrm{DMD}}
  \;=\;
  ({\sf U}_r^{\mathrm{DMD}})^{\!\top} \, {\sf Y}_d^{(d,N)} \, {\sf V}_r^{\mathrm{DMD}} \, \Sigma_r^{-1}.
\end{align}
Let ${\sf A}^{\mathrm{DMD}} {\sf W} = {\sf W} \Lambda$ be its eigendecomposition with
$\Lambda=\mathrm{diag}(\lambda^{\mathrm{DMD}}_{1},\ldots,\lambda^{\mathrm{DMD}}_{r})$ and ${\sf W}=[\vct{w}_1,\ldots,\vct{w}_r]$.
The DMD modes in the data space are
\begin{align}
  \Phi
  \;=\;
  {\sf Y}_d^{(d,N)} \, {\sf V}_r^{\mathrm{DMD}} \, \Sigma_r^{-1} \, {\sf W}
  \;=\;
  [\,\boldsymbol{\phi}_1,\ldots,\boldsymbol{\phi}_r\,].
\end{align}
Continuous-time exponents are obtained by
\begin{align}
  \nu^{\mathrm{DMD}}_{i}
  \;=\;
  \frac{1}{\Delta t}\log\!\big(\lambda^{\mathrm{DMD}}_{i}\big)
  \;=\;
  \alpha^{\mathrm{DMD}}_{i} + \jj \,\omega^{\mathrm{DMD}}_{i},
\end{align}
and the associated period is
\begin{align}
  T_i^{\mathrm{DMD}}
  \;=\;
  \frac{2\pi}{|\omega^{\mathrm{DMD}}_{i}|},
  \qquad \mathrm{if}\,\,\omega_i^{\mathrm{DMD}} \neq 0.
\end{align}
The quantity $T_i^{\mathrm{DMD}}$ indicates the characteristic time scale of (possibly oscillatory) temporal patterns in $y[n]$.
When visualizing $T_i^{\mathrm{DMD}}$, we omit strictly non-oscillatory modes with $|\omega^{\mathrm{DMD}}_{i}|=0$.

To quantify the dominance of each DMD mode, we evaluate its global projection contribution over the Hankel snapshots.
For the $i$-th DMD mode $\boldsymbol{\phi}_i \in \mathbb{C}^{d^\ast}$, the orthogonal projector onto $\mathrm{span}\{\boldsymbol{\phi}_i\}$ is
\begin{equation}
  {\sf P}_i
  \;:=\;
  \frac{\boldsymbol{\phi}_i \, \boldsymbol{\phi}_i^{\sf H}}{\boldsymbol{\phi}_i^{\sf H}\boldsymbol{\phi}_i}
  \;\in\; \mathbb{C}^{d^\ast \times d^\ast},
  \label{eq:dmd_proj_op}
\end{equation}
where ${\sf H}$ denotes the Hermitian transpose.
Applying ${\sf P}_i$ to the snapshot matrix yields the portion explained by the $i$-th mode:
\begin{equation}
  {\sf X}_i^{\mathrm{proj}}
  \;:=\;
  {\sf P}_i \, {\sf X}_d^{(d,N)}
  \;\in\; \mathbb{C}^{d^\ast \times (N-d^\ast)}.
  \label{eq:dmd_proj_part}
\end{equation}
We define the (global) contribution ratio of mode $i$ by the Frobenius-norm energy ratio
\begin{equation}
  c_i^{\mathrm{DMD}}
  \;:=\;
  \frac{\big\|{\sf X}_i^{\mathrm{proj}}\big\|_{\mathrm F}^{2}}
       {\big\|{\sf X}_d^{(d,N)}\big\|_{\mathrm F}^{2}}
  \;=\;
  \frac{\big\|{\sf P}_i \, {\sf X}_d^{(d,N)}\big\|_{\mathrm F}^{2}}
       {\big\|{\sf X}_d^{(d,N)}\big\|_{\mathrm F}^{2}}.
  \label{eq:dmd_contrib}
\end{equation}

In the following, we visualize $T_i^{\mathrm{DMD}}$ versus $c_i^{\mathrm{DMD}}$ to interpret the time scales of dominant patterns extracted by Delay-DMD.

\vspace{0.5em}
\subsection{Baseline results}
\label{subsec:baseline_results}

\subsubsection{SSA}

\begin{figure}[!t]
  \begin{center}
    \includegraphics[width=\hsize]{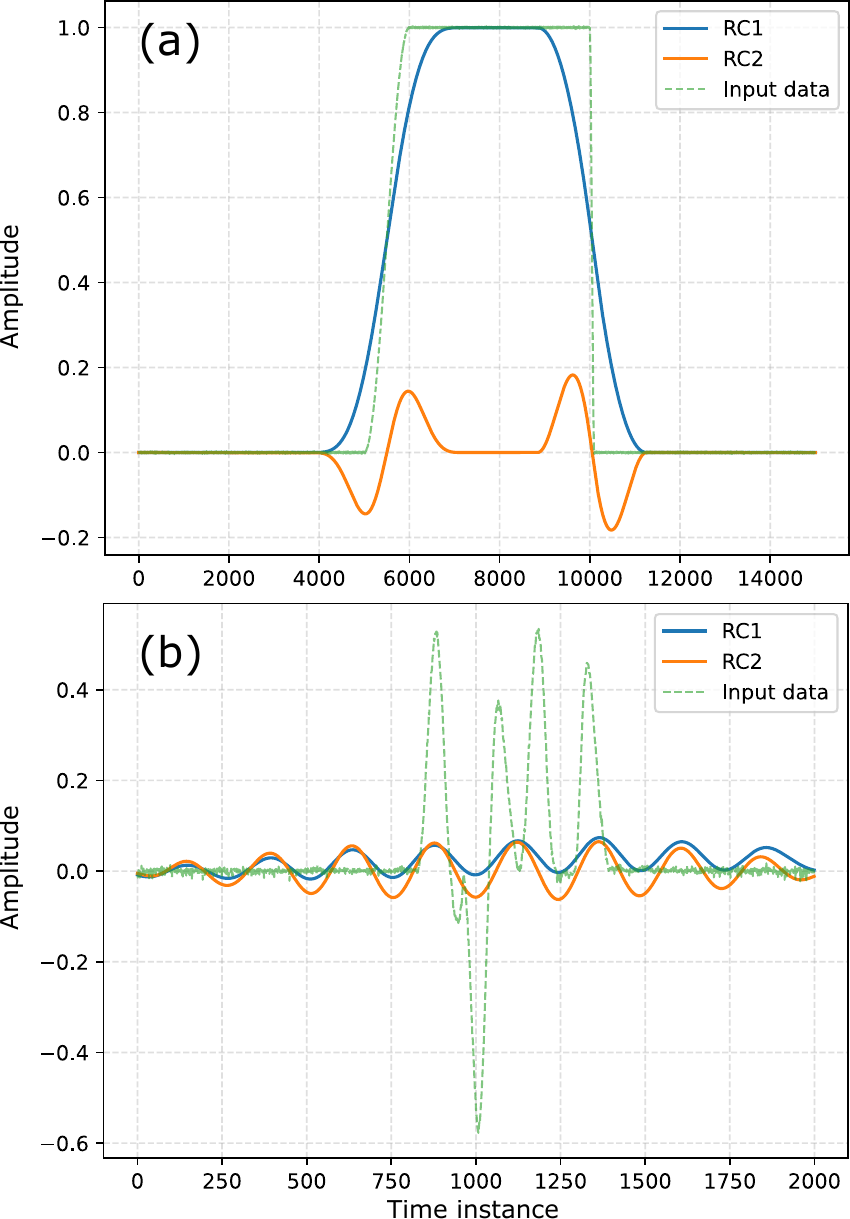}
    \caption{SSA results for (a) toy model data and (b) turret punch data.
    The blue and orange lines show reconstructed signals $\mathrm{RC}_{1}[n]$ and $\mathrm{RC}_{2}[n]$, respectively.
    The dashed green line shows the input data (toy model or turret punch).}
    \label{fig:SSA_reconstruct}
  \end{center}
\end{figure}

The decomposed waveforms obtained by SSA are shown in \figref{fig:SSA_reconstruct}(a) for the toy model and \figref{fig:SSA_reconstruct}(b) for the turret punch data,
where the blue and orange lines show reconstructed signals $\mathrm{RC}_{1}[n]$ and $\mathrm{RC}_{2}[n]$, respectively, and the dashed green line shows the input data (toy model or turret punch).
For Fig.~\ref{fig:SSA_reconstruct}(a), SSA largely reproduces the original input in the leading $\mathrm{RC}_{1}[n]$ and fails to produce a clean separation into distinct pulsive sub-patterns.
The second component $\mathrm{RC}_{2}[n]$ contains isolated sinusoidal segments that do not correspond to the designed pulsive structure.
For Fig.~\ref{fig:SSA_reconstruct}(b), both $\mathrm{RC}_{1}[n]$ and $\mathrm{RC}_{2}[n]$ indicate nearly steady sinusoidal curves, so that the transient patterns in the original input are not represented.
Thus, in both cases, SSA does not separate the original signal into interpretable pulsive components.

\vspace{0.5em}
\subsubsection{Delay-DMD}

\begin{figure}[!t]
  \begin{center}
    \includegraphics[width=\hsize]{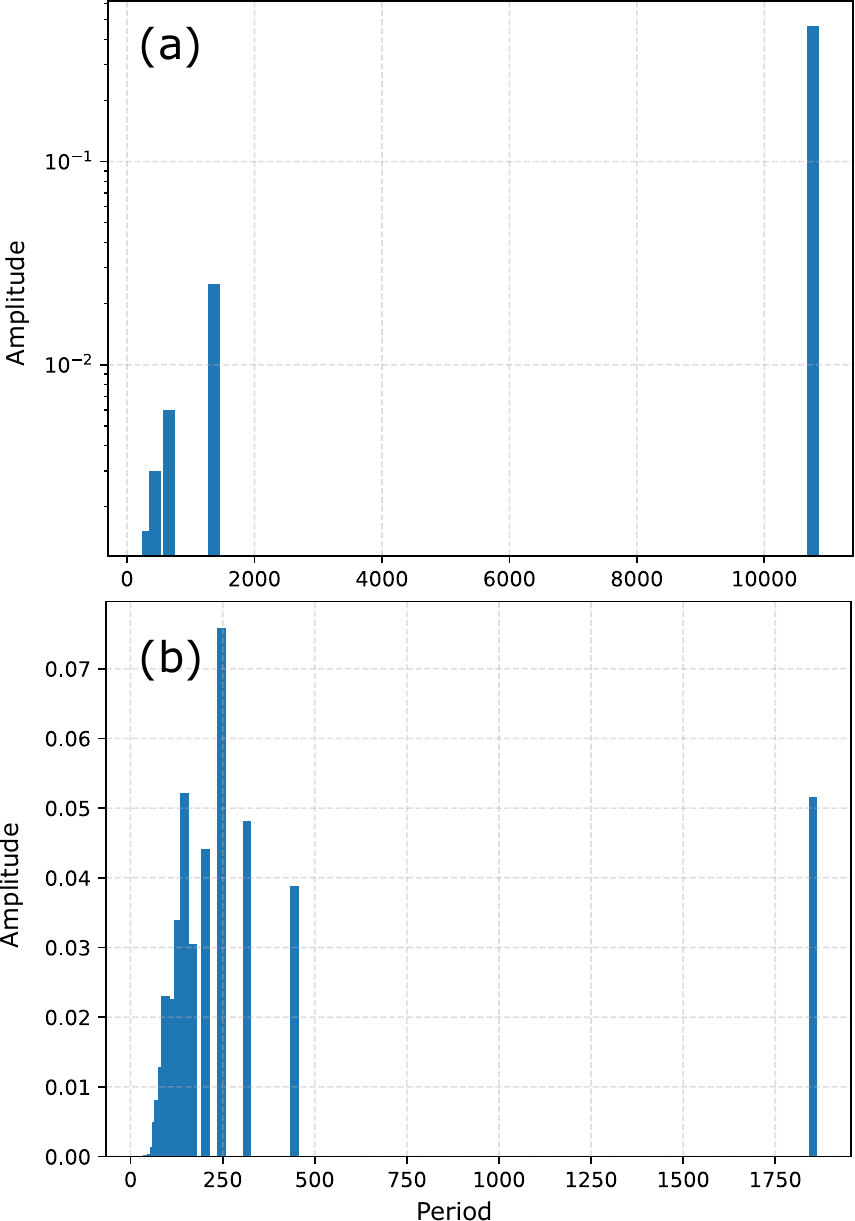}
    \caption{DMD results for (a) toy model data and (b) turret punch data:
    period $T_i^{\mathrm{DMD}}$ versus contribution ratio $c_i^{\mathrm{DMD}}$ for each mode.}
    \label{fig:DMD_energy}
  \end{center}
\end{figure}

The period $T_i^{\mathrm{DMD}}$ versus contribution ratio $c_i^{\mathrm{DMD}}$ for each mode obtained by Delay-DMD are shown in \figref{fig:DMD_energy}(a) for the toy model and \figref{fig:DMD_energy}(b) for the turret punch data.

In Fig.~\ref{fig:DMD_energy}(a), most DMD modes have real (non-oscillatory) eigenvalues; therefore, only five pairs of complex-conjugate modes are plotted.
The contribution ratio $c_i^{\mathrm{DMD}}$ takes the largest value at $T_i^{\mathrm{DMD}}=10770.77$, which corresponds to the total data length $N=12000$.
The second dominant scale, $T_i^{\mathrm{DMD}}=1369.24$, might correspond to $\tau_0=1000$ of the first half-sine segment in the toy model data.
The smallest $T_i^{\mathrm{DMD}}$ among the oscillatory modes is $323.63$, which is significantly larger than $\tau_1=100$.
This indicates that Delay-DMD 
fails to isolate short-lived transient segments in the toy model.

In Fig.~\ref{fig:DMD_energy}(b), the contribution ratio $c_i^{\mathrm{DMD}}$ takes the largest value at $T_i^{\mathrm{DMD}}=1852.31$, which corresponds to the total data length $N=2000$.
The second-largest $T_i^{\mathrm{DMD}}$ is $445.63$, which is smaller than the distance $L_{0,8}$ between the locations $L_0$ and $L_8$.
This suggests that Delay-DMD does not recover the overall pulsive pattern spanning $L_0$ to $L_8$.
Moreover, for $T_i^{\mathrm{DMD}} < 200$, the spectrum exhibits no significant peaks, implying that Delay-DMD fails to distinguish short-term rising or falling segments contained in the pulsive oscillation of the turret punch data.

\subsection{Comparative discussion}

The comparative analysis highlights fundamental limitations of the Hankel-based baselines in the present setting.
SSA, although effective for decomposing long signals into oscillatory components, lacks a principled,
unified way to quantify the time scale of each reconstructed component.
As shown in \secref{subsec:baseline_results}, SSA tends to reproduce the original signal in the leading component and to generate residual sinusoidal artifacts in subsequent ones.
Such decompositions fail to capture localized one-shot pulsive events in both the toy model and the turret data.

Delay-DMD, on the other hand, explicitly assigns oscillatory time scales via eigen-analysis.
However, the contribution spectra in \figref{fig:DMD_energy} show that the dominant modes correspond either to the full data length or to broad oscillatory envelopes.
Short transient scales---such as the falling segment of the toy waveform or localized rises in the turret sequence---are not represented by single modes; multiple oscillatory modes must be combined to approximate such non-stationary structures, which undermines physical interpretability.

In contrast, the proposed DEF method is explicitly designed to isolate one-shot pulsive structures.
By aligning the embedding with the event length $d^\ast$, DEF directly captures the dominant transient scales without requiring repeated instances or stationary oscillations.
Across both datasets, DEF identifies the correct pulsive scales where SSA fails to separate and Delay-DMD over-smooths.
This capability is particularly valuable in industrial anomaly detection scenarios, 
where available data consist of sparse, one-shot transients rather than long stationary time series.

Overall, while SSA and Delay-DMD are useful generic Hankel-based decompositions, they do not, by construction, target non-repeating pulsive patterns.
DEF provides a compact and interpretable representation of such dynamics and, in the present experiments, uniquely recovers the dominant pulsive time scales that align with the empirical structures reported in \secref{sec:numerical} and \secref{sec:demo}.

\section{Conclusions}
\label{sec:concusions}

In this paper, we presented an alternative approach to analyzing temporal patterns by focusing on their duration.
We studied time-series data $\{y(t)\}$ that resulted from a state space system controlled by a piecewise constant control input $\vct{u}(t)$.
By identifying a linear dynamical model of subsequences (delay coordinates) and assuming that its discrete spectrum can describe the target signal $y(t)$, we developed a framework called Data-driven Exponential Framing (DEF).
DEF allows us to capture exponential decay rates associated with dominant temporal patterns based on the eigen-decomposition of the identified model.
We demonstrated that DEF can identify two distinct lengths (or timescales) of subsequences in numerical data obtained from a harmonic oscillator-based model.
We also applied DEF to pulsive current data collected from a turret punch and showed that the timescales derived by DEF can explain any combination of fundamental curves of the data.
Futhermore, comparative analysis with SSA and DMD clearly shows the uniqueness of DEF as a compact and interpretable representation non-repeating pulsive patterns.
Therefore, we argue that DEF can characterize temporal patterns inherent in time series data by their duration (timescales).

DEF can be applied to a wide range of data that can be described by their
discrete spectrum,
as it requires only the two design parameters ($L$ and $\sigma^2$) and does not require any prior information on the systems.
While our main experiments considered piecewise constant exogenous signals,
the formulation naturally generalizes by augmenting input lags in the delay vector when inputs exhibit intra-plateau dynamics (see Appendix A), preserving the same estimation pipeline.
In addition, the framework is not restricted to single-output settings; extending it to multi-output problems is straightforward and may further broaden its applicability.

Finally, we emphasize the practicality of DEF.
It enables extract fundamental temporal patterns from data, which can be reused as search queries to explore larger datasets.
For system modeling and control,
the bases $\tilde{\vct{V}}_i$ of DEF compactly represent pulse-driven dynamics, providing an efficient means for dimensionality reduction.
Beyond these points, the DEF outputs serve directly in downstream analysis and design:
(i) as \emph{pattern-level descriptors} that validate physics-based models by comparing predicted versus observed time scales;
(ii) as compact targets for controller tuning and setpoint scheduling, where desired temporal responses (rise/fall durations) are specified and checked against $\{\tilde{T}_i\}$ and $\{\tilde{A}_{i,d^*}[n]\}$;
and (iii) as search/indexing keys to retrieve similar episodes across different runs, lots, or machines.
Importantly, DEF is complementary to known physics in manufacturing:
it provides a data-grounded summary of dominant pulse-like phenomena without requiring a full parametric model a priori,
while still constraining and informing such models by exposing the critical timescales that must be reproduced.

\begin{acknowledgment}
The authors appreciate Mr.~Yoshinori Mochizuki, Mr.~Ippei Takasawa, Mr.~Takuma Nishimura, Mr.~Atsushi Otake, and Dr.~Kenta Deguchi (Hitachi, Ltd.) for valuable discussion and provision of the measurement data.
\end{acknowledgment}

%
% ...

\bibliographystyle{asmems4}

\bibliography{self}

\section*{Appendix A: Relaxation of the piecewise-constant input assumption}

In Sec.~2 we introduced DEF under the assumption that the exogenous input $\vct{u}(t)$ is piecewise constant (PWC). 
Here, we relax this restriction by augmenting the delay coordinates with input lags, thereby extending the formulation to general forced systems without altering the overall pipeline.

Consider extending \eqref{eq:delay_model} to incorporate input lags:
\begin{align}
  \frac{\dd\vct{x}_d(t)}{\dd t}
  = [{\sf A}_d, {\sf B}_d]
  \begin{bmatrix}
  \vct{x}_d(t)\\
  \vct{u}_d(t)
  \end{bmatrix},
  \label{eq:delay_model_ext}
\end{align}
where ${\sf B}_d \in \mathbb{R}^{d \times dm}$ and 
$\vct{u}_d(t) \in \mathbb{R}^{dm}$ denotes the delay-coordinate vector of the input with sampling interval $\Delta t$:
\begin{align}
  \vct{u}_d(t) =
  \begin{bmatrix}
  \vct{u}(t)\\
  \vct{u}(t-\Delta t)\\
  \vdots\\
  \vct{u}(t-(d-1)\Delta t)
  \end{bmatrix}.
\end{align}
Including input delays provides a practical embedding for driven or controlled systems. 
For the PWC case treated in Sec.~2, the term ${\sf B}_d\vct{u}_d(t)$ becomes constant within each plateau and is absorbed into the intercept, leaving the identified ${\sf A}_d$ unchanged. 
Thus, \eqref{eq:delay_model_ext} can be viewed as a natural extension of the state equation in Sec.~2.

Once ${\sf A}_d$ and ${\sf B}_d$ are identified from data, the DEF decomposition extends in a straightforward way. 
For instance, the $1$-step prediction becomes
\begin{align}
\vct{x}_d(t_0 + \Delta t) - {\sf B}_d\vct{u}_d(t_0) 
&= \exp({\sf A}_d \Delta t)\, \vct{x}_d(t_0) \nonumber\\
&= \sum_{i=1}^{d} a_{i,d}(t_0)\, \exp(\lambda_i \Delta t)\, \vct{V}_i,
\label{eq:seq_evol_ext}
\end{align}
which is the forced analogue of \eqref{eq:seq_evol}. 
Accordingly, the subsequence expansion \eqref{eq:delay_expansion} is modified by subtracting the input contribution: for $n = 0, 1, \ldots, d-1$,
\begin{align}
  y(t-(n-1)\Delta t) - b_{n,d}(t) \;\simeq\; 
  \sum_{i=1}^{d} \tilde{a}_{i,d}(t)\, \exp(-\nu_i n\Delta t),
  \label{eq:delay_expansion_ext}
\end{align}
where $b_{n,d}(t)$ is the $n$-th component of ${\sf B}_d\vct{u}_d(t)$. 
Thus, the amplitude estimation in \eqref{eq:fit} becomes
\begin{align}
  \tilde{a}_{i,d}(t) \;\simeq\; 
  \sum_{n=0}^{d-1} \big(y(t-(n-1)\Delta t) - b_{n,d}(t)\big)\,
  \exp(\nu_i n \Delta t).
  \label{eq:fit_ext}
\end{align}
This adjustment shows that the input effect is explicitly removed prior to estimating $\tilde{a}_{i,d}(t)$.

In terms of system identification, both ${\sf A}_d$ and ${\sf B}_d$ are obtained by minimizing a modified regression error:
\begin{align}
  E_d = \left\|
    {\sf Y}_d^{(d,N)} - \tilde{\sf A}_d {\sf X}_d^{(d,N)} 
    - \tilde{\sf B}_d {\sf U}_d^{(d,N)}
  \right\|_\mathrm{F}^2,
  \label{eq:reg_error_forced}
\end{align}
where ${\sf U}_d^{(M_1,M_2)} = [\vct{u}_d(M_1-1), \ldots, \vct{u}_d(M_2-2)]$. 
Similarly, the AIC criterion in \eqref{eq:AIC} extends to
\begin{align}
  \mathrm{AIC}(d) =& 
  \frac{\left\|
  {\sf Y}_d^{(d+L-1,N)} - {\sf C}_d {\sf Z}_{d,L}
  \right\|^2}{N-d-L}
  + \frac{2d}{N-d-L}\sigma^2,
  \label{eq:AIC_forced}
\end{align}
with 
\begin{align}
  {\sf Z}_{d,L} := \tilde{\sf A}_d^L {\sf X}_d^{(d,N-L+1)} 
  + \sum_{i=1}^L \tilde{\sf A}_d^{L-i}\tilde{\sf B}_d\, {\sf U}_d^{(d+i-1,N-L+i)}.
\end{align}
Finally, the amplitude at time $n$ is evaluated as
\begin{align}
  \tilde{A}_{i,d^*}[n] = 
  \big(\vct{x}_d[n+1] - {\sf B}_d\vct{u}_d[n]\big)^\top \tilde{\vct{V}}_i.
\end{align}

In summary, the extension above preserves the structure of the DEF pipeline presented in Sec.~2. 
The only difference is the explicit subtraction of the input contribution through the matrices ${\sf B}_d$ and ${\sf U}_d$, while all subsequent steps (eigendecomposition, extraction of $\nu_i$, and amplitude evaluation) remain identical. 
This guarantees that the methodology is not restricted to PWC inputs and can be seamlessly applied to general forced systems.

\section*{Appendix B: Effect of the horizon length $L$ on the shape of the information criterion}

To illustrate how the choice of horizon length $L$ affects the information criterion,
we compared three representative settings: $L=1$, $L=5$, and $L=20$.
Since the absolute value of $\mathrm{AIC}(d)$ varies with $L$, we normalized each curve by subtracting its minimum value, i.e.,
\begin{equation}
  \mathrm{AIC}_{\mathrm{rel}}(d;L) 
  = \mathrm{AIC}(d;L) - \min_{d} \mathrm{AIC}(d;L),
\end{equation}
so that all curves are aligned at zero.
This normalization allows us to compare the shape and the consistency of discriminative ability across different $L$.

\begin{figure}[!t]
  \begin{center}
    \includegraphics[width=\hsize]{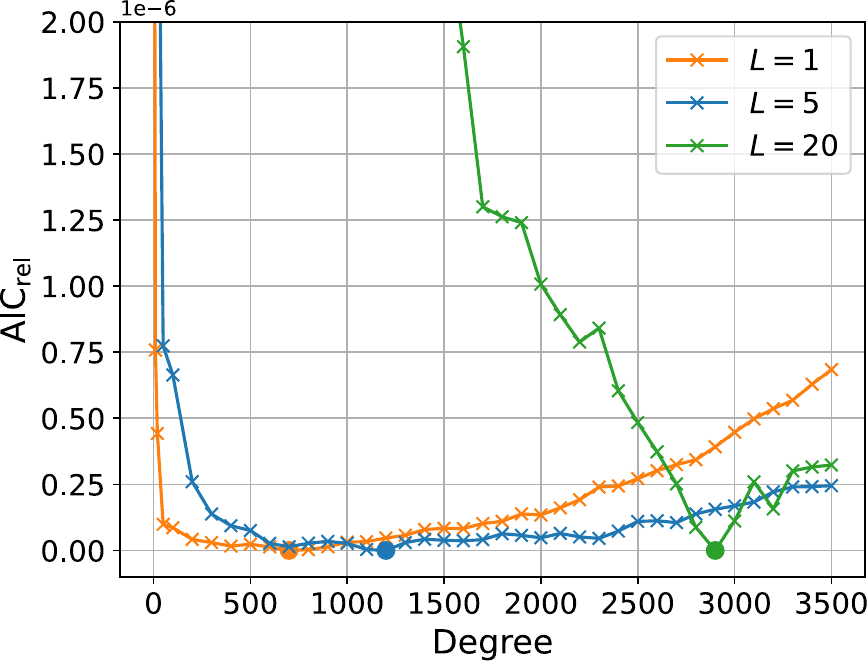}
    \caption{$\mathrm{AIC}_{\mathrm{rel}}(d;L)$ for $L = 1, 5 ,20$.}
    \label{fig:AIC_comparison}
  \end{center}
\end{figure}

Figure~\ref{fig:AIC_comparison} shows the $\mathrm{AIC}_{\mathrm{rel}}(d;L)$ curves for $L=1$, $L=5$, and $L=20$.
For $L=1$, the curve is convex but exhibits a steep rise for larger $d$, which gives the appearance of a trivial or ``already-known'' solution.
This reduces the practical informativeness of the criterion.
For $L=5$, the curve is convex with a gradual rise, striking a good balance between discriminative sharpness and robustness across model orders.
For $L=20$, the curve tends to slope downwards, becoming susceptible to noise accumulation, which undermines the reliability of model order comparison.
These observations support our heuristic choice of $L=5$ in the main experiments, as it achieves a balance between consistency and discriminative effectiveness.

\section*{Appendix C: Peak extraction algorithm}

For completeness, we provide here the definitions and explanations underlying Algorithm~2 (Peak extraction).  
Let $\tilde{T}_i$ denote the estimated period corresponding to the $i$-th delay component,
and let $\tilde{A}_{i,d^*}[n]$ be the associated amplitude spectrum at time index $n$ with embedding order $d^*$.  
The goal is to extract dominant peaks in the $(\tilde{T}_i, \tilde{A}_{i,d^*})$ spectrum.

The scoring criterion combines two factors:  
(i) the peak prominence $p_i$, defined as the excess amplitude above a local baseline computed within a sliding window of log-period width $w$, and  
(ii) the isolation score $s_i$, which measures how well the candidate stands out relative to the local background energy $e_i$ in its neighborhood.  
A small stabilizer $\epsilon$ is included to ensure numerical robustness when the background energy is close to zero.

The final dominance score is given by $q_i = p_i \cdot s_i$, balancing absolute height with relative isolation.  
Peaks with $q_i$ above a threshold $\Theta$ are considered significant, and among them, up to $K$ top-ranked peaks are retained for reporting.  
These parameters ($w, \epsilon, \Theta, K$) were fixed at default values in our experiments, but can be tuned without altering the qualitative outcome.

This procedure corresponds to the heuristic visual identification employed in the main experiments, while the algorithm provides a reproducible formalization of the same criterion.
\begin{algorithm}[tb]
  \caption{Peak extraction on time-scale spectrum $(\tilde{T}_i,\,\tilde{A}_{i,d^*}[n])$ at a fixed time $n$}
\KwIn{Pairs $\{(\tilde{T}_i,\,\tilde{A}_{i,d^*}[n])\}_{i=1}^{d^*}$ at a fixed $n$}
\KwOut{Reported dominant peaks $\mathcal{P}_{\mathrm{rep}}(n)$}

$a_i \leftarrow |\tilde{A}_{i,d^*}[n]|,\quad x_i \leftarrow \log \tilde{T}_i$\;
Sort $\{(x_i,a_i)\}$ in ascending $x_i$\;

$b_i \leftarrow \mathrm{running\_median}(a_i;\ w)$\;
$p_i \leftarrow \max(0,\, a_i - b_i)$\;

\For{$i=1$ \KwTo $d^*$}{
  $\mathcal{N}_i \leftarrow \{\, j\ne i\ |\ w/2 \le |x_j-x_i| \le w \,\}$\;
  $e_i \leftarrow \sqrt{\tfrac{1}{|\mathcal{N}_i|}\sum_{j\in\mathcal{N}_i} a_j^2}$\;
  $s_i \leftarrow \dfrac{p_i}{\epsilon + e_i}$\;
}
$q_i \leftarrow p_i \cdot s_i$\;

$\mathcal{C} \leftarrow \emptyset$\;
\For{$i=1$ \KwTo $d^*$}{
  \If{$q_i$ is the maximum within $\{j:\ |x_j-x_i|\le w\}$}{
    Add $(i,\ \tilde{T}_i,\ a_i,\ q_i)$ to $\mathcal{C}$\;
  }
}

\textbf{rank} $\mathcal{C}$ by $q_i$ (descending)\;
$\mathcal{P}_{\mathrm{all}}(n) \leftarrow \{\, (i,\tilde{T}_i) \in \mathcal{C}\ |\ q_i \ge \theta \,\}$\;
$\mathcal{P}_{\mathrm{rep}}(n) \leftarrow$ top-$K$ entries of $\mathcal{C}$ by $q_i$\;
\end{algorithm}

\end{document}